\documentstyle[twocolumn,prl,aps]{revtex}

\begin{document}
\draft
\preprint{}
\title{Critical Properties in Photoemmision Spectra \\           
        for One Dimensional Orbitally Degenerate Mott Insulator}
\author{ Tatsuya Fujii, Yasumasa Tsukamoto and Norio Kawakami } 
\address{Department of Applied Physics,
Osaka University, Suita, Osaka 565, Japan} 
\date{\today}
\maketitle
\begin{abstract}
Critical properties in photoemission spectra for 
the one-dimensional Mott insulator with orbital degeneracy  
are studied by exploiting  the integrable {\it t-J} model,
which is a supersymmetric generalization of the 
SU($n$) degenerate spin model. We discuss
the critical properties for  
the holon dispersion as well as the spinon dispersions, by applying
the conformal field theory analysis
to the exact finite-size energy spectrum.
We study the effect of orbital-splitting 
on the spectra by evaluating the momentum-dependent 
critical exponents. 
\end{abstract}
\section{Introduction}
The recent experiments on the angular resolved 
photoemission for one-dimensional (1D) correlated
electron systems\cite{ex1,ex2,ex3,ex4,ex5,ex6,ex7} 
have revealed some striking  properties inherent in 
1D systems, such as the spin-charge separation.   
In particular, the photoemission experiments\cite{ex5,ex6,ex7} for the 
Mott insulator compound SrCuO$_{2}$, Sr$_{2}$CuO$_{3}$ and NaV$_{2}$O$_{5}$ 
have clarified characteristic properties of 
a single hole doped in the 1D Mott insulator.
So far, extensive theoretical studies on spectral 
functions for 1D correlated systems
have been done numerically 
\cite{shiba,preuss,haas,ppenc,maekawa,loren,mila,zacher}
and analytically. \cite{S&P1,meden,voit2,S&P2,voit,nagaosa,fujii}
Among others, it has been recently shown
\cite{S&P1,S&P2,voit} that the dynamical
spectral functions for the 1D Mott insulator 
still show the  power-law singularity at threshold
energies, characterizing the critical properties of a 
doped hole in the 1D Mott insulator. For example, 
Sorella and Parola\cite{S&P2} have addressed this problem 
based on the  conformal 
field theory (CFT) analysis of the supersymmetric
{\it t-J } model. Also, such an analysis has been
extended to the anisotropic {\it t-J} model
with either the spin gap or the charge gap.\cite{fujii}

In some compounds of the Mott insulator,
such as Mn oxides, it is known that the orbital degrees of freedom
as well as the spin degrees of freedom
play an important role.\cite{mott} These hot topics on the Mott
insulator have stimulated the theoretical studies on the  
orbitally degenerate Hubbard model and the related
correlated electron systems. As a first step  
to explore such orbital effects more precisely, the 1D version of 
the degenerate Hubbard model in the strong coupling regime
has been extensively studied.\cite{ueda,zhang,troy}

Motivated by the above investigations, we study the
critical properties in 
photoemission spectra for the 1D Mott insulator
with orbital degeneracy.
For this purpose, we consider the 1D SU($n$)
spin model as a Mott insulator in the strong coupling 
regime. Doping a hole into the model is  
incorporated by exploiting  the integrable {\it t-J} model,
which is a supersymmetric generalization of the 
SU($n$) spin model.\cite{suther} We shall discuss
the critical properties in photoemission spectra 
for the holon as well as spinon dispersions, 
based on the CFT analysis of the above integrable model.

This paper is organized as follows. 
In {\S} 2 
we introduce the model, and present the calculational detail
which is necessary to study the low-energy critical
properties. Our main strategy is the finite-size scaling 
technique applied to the energy spectrum, in which 
we should properly
take into account a final-state interaction between the
massless spinons and the massive  holon created by
photoemission.  Then in {\S} 3 we derive the spectral functions 
and their critical exponents based on the CFT analysis. 
We discuss photoemission spectra 
for the massive holon dispersion for SU($n$) symmetric case,
and then applying a similar technique, we also study
a power-law singularity for the $(n-1)$ spinon dispersions.
We discuss the effect of orbital splitting 
by exploiting the SU(4) spin model with 
two-band structure, for which we obtain the 
momentum-dependent critical exponents for the 
photoemission spectra. Brief summary is given in {\S} 4.

\section{Conformal Properties of Orbitally Degenerate 
Mott Insulator with a Hole}

\subsection{Model and basic equations}
Let us start with the ordinary Hubbard model
with on-site Coulomb interaction and Hund coupling, 
which possesses $L$-orbitals and spins $\sigma=\uparrow,\downarrow$. 
In the case of the strong on-site Coulomb interaction in the
insulating phase, the model is generally 
reduced to the Kugel-Khomskii type model.\cite{kugel} If we neglect  
the Hund coupling for simplicity, the symmetry 
of the system is enhanced and the resulting model 
becomes the SU($2L$) exchange model,
for which $2L$ indices specify both
of the orbital and spin degrees of freedom.\cite{ueda,zhang}
We study this  exchange model as the 
spin and orbital parts of the Mott insulator. Doping  a hole 
into the exchange model induces a kinetic term, which
should be taken into account to discuss the photoemission spectrum.
To this end, we use the supersymmetric {\it t-J}
model for which the electron hopping $t$ is assumed to be 
equal to the exchange coupling $J$.  The advantage 
to use this model is that it allows us to treat various quantities 
exactly.  Let us now write down the Hamiltonian for the
multi-orbital {\it t-J} model, 
\begin{eqnarray}
{\cal H}
%
&=&-t{\sum_{i}}{\sum_{\sigma,m}}
(c^{m\dagger}_{i,\sigma}c^{m}_{i+1,\sigma}
                          +c^{m\dagger}_{i+1,\sigma}c^{m}_{i,\sigma}) \cr
&+&J{\sum_{i}}{\sum_{\sigma,{\sigma}^{'}}}{\sum_{m,m^{'}}}
                 (c^{m\dagger}_{i,\sigma}c^{m^{'}}_{i,{\sigma}^{'}}
                  c^{m^{'}\dagger}_{i+1,{\sigma}^{'}}c^{m}_{i+1,\sigma}- n_{i}n_{i+1})\cr
&-&\sum_{m}{\sum_{i}}{\Delta_{m}}(n^{(m)}_{i}-n^{(m+1)}_{i}),
\label{eqn:su-1}
\end{eqnarray}
where $c^{m\dagger}_{i,\sigma}$ is the creation operator for electrons. 
It is implicitly assumed that multiple occupation of 
each site is forbidden. 
Each orbital, which is labeled by $m$($=1,2, \cdots, L$)
and is energetically separated 
from each other by the orbital-splitting $\Delta_{m}$, has 
two spin states  ($\sigma=\uparrow,\downarrow$). 
At the supersymmetric point $t = J$
which we are interested in, the model is 
exactly solvable.\cite{suther,schlott2} 

In the following, we exactly analyze the low-energy properties 
of the model in the insulating state (one electron for every site)
with {\it one hole}, since one electron is emitted from the 
system in photoemission experiments. To this end, we first write down 
the relevant Bethe equations for the model (\ref{eqn:su-1}), 
which describe excitations when {\it one electron is emitted 
from the insulating system}. For convenience, we introduce 
the number $n=2L$ and the index $\mu$($=1,\cdots,n-1$)
which specifies the spin as well as orbital excitations
on equal footing. 
The Bethe equations consist of $(n-1)$ kinds of rapidities 
$\lambda_{\alpha}^{(\mu)}$ for spin and orbital 
degrees of freedom,\cite{suther} 
\begin{eqnarray}
  - N \Theta_{2}(\lambda_{\alpha}^{(\mu)}) \delta_{\mu 1} 
   &+& 2\pi I_{\alpha}^{(\mu)}   
  = \sum_{\nu=1}^{n-1} \sum_{\beta=1}^{M_{\mu}} 
    \Theta_{\mu \nu}(\lambda_{\alpha}^{(\mu)} - \lambda_{\beta}^{(\nu)}) \cr
    &+&
    \Theta_{2}(\lambda_{\alpha}^{(\mu)} - q)\delta_{\mu n-1}
    \hspace{4mm} 1\leq \alpha \leq M_{\mu}, \cr
    &\hspace*{1mm}&
\label{eqn:su-2}
\end{eqnarray}
where $\Theta_{\mu\nu}=-\Theta_{1}\delta_{\mu \nu}
  +\Theta_{2}(\delta_{\mu \nu+1}+\delta_{\mu+1 \nu}),$ 
and $\Theta_{n}(\lambda)=2 \tan^{-1} (n\lambda)$. 
Here we have introduced the quantity, 
$M_{\mu}=\sum_{\gamma=\mu+1}^{n+1} N_{\gamma}$, 
in (\ref{eqn:su-2}), where $N_{\gamma}$ is the number of particles 
which belong to the type  $\gamma$ ($\gamma=1,\cdots ,n+1$). 

It is to be noticed here that there exists 
{\it an impurity term} 
$\Theta_{2}(\lambda_{\alpha}^{(\mu)}-q) \delta_{\mu n-1}$
in the above Bethe equations.  This is the term
which indeed describes a scattering 
due to the  hole created in photoemission, where 
a charge rapidity $q$ specifies the momentum  
of the created hole. In this sense, this scattering term 
due to the created hole can be regarded as a final-state interaction
caused by the photoemission.  
For simplicity, in this paper 
we shall refer to $(n-1)$ massless excitations
as spinons while the above charge excitation as holon.
Needless to say, the holon excitation is massive because
of the existence of the Mott-Hubbard gap.
We will see that although massive holon excitation 
does not directly enter in the low energy physics, 
it controls the critical behavior of massless spin excitations 
via a final-state interaction. 

Finally we note that each quantum number $I_{\alpha}^{(\mu)}$
is subject to the selection rule (gluing condition for electrons),
\begin{eqnarray}
  I_{\alpha}^{(\mu)}=-\frac{1}{2}(M_{\mu -1}-M_{\mu}+M_{\mu +1})
                 +M_{\mu+1} \delta_{\mu 1} \hspace{2mm} {\rm mod}1,\cr
\label{eqn:su-3}
\end{eqnarray}
which plays an essential role when we read 
the critical exponents from the finite-size spectrum.

\subsection{Finite-size corrections}

In order to apply the methods developed in  CFT,\cite{bpz} 
we now evaluate the finite-size corrections to the ground-state 
energy and the  excitation energy.\cite{cardy,affleck}
The key point in our case is to take into account a final-state 
interaction properly, i.e. the scattering phase shift due to 
the massive holon. Since the calculation of the spectrum is performed 
in a standard way,\cite{woy,S&P2}
 we briefly summarize the outline of the 
calculation in the following.  We also mention that the 
present calculation has a close relationship to that for 
1D solvable systems with a impurity or 
boundaries.\cite{e1,e2,e3,e4,e5,e6,e7}

The total energy and the total momentum are given 
in terms of the rapidities, 
\begin{eqnarray}
E=\sum_{\alpha=1}^{n} \sum_{j=1}^{M_{\alpha}}
    \varepsilon^{0}_{\alpha}(\lambda_{j}^{\alpha}), \hspace{2mm}
P=\sum_{\alpha=1}^{n} \sum_{j=1}^{M_{\alpha}}
    p^{0}_{\alpha}(\lambda_{j}^{\alpha}).
\label{eqn:su-4}
\end{eqnarray}
According to the Euler-Maclaurin formula, we can obtain 
the finite-size corrections to the ground-state energy
in a standard form,  
\begin{eqnarray}
   E_{0}=N \epsilon_{0} - \sum^{n-1}_{\alpha=1} 
\frac{\pi v_{\alpha}}{6N},
\label{eqn:su-5}
\end{eqnarray}
where $N \epsilon_{0}$ is the ground-state energy 
in the thermodynamic limit. 
Here $v_{\alpha}$ is the velocity of massless spinons 
labeled by $\alpha$, 
\begin{eqnarray}
v_{\alpha}=\frac{\varepsilon^{'}_{\alpha}(\lambda_{\alpha})}
                {2\pi \sigma_{\infty \alpha}(\lambda_{\alpha})} 
                  \left |_{\lambda_{\alpha}=\lambda_{\alpha}^{0}} 
\right. ,
\label{eqn:su-6}
\end{eqnarray}
where 
\begin{eqnarray}
   \varepsilon_{\alpha}(\lambda_{\alpha}|\lambda^{\pm}_{\nu})
&=&
   \varepsilon^{0}_{\alpha}(\lambda_{\alpha}) \cr
&+& \sum_{\gamma =1}^{n-1} \int_{\lambda^{-}_{\gamma}}^{\lambda^{+}_{\gamma}}
       K_{\alpha \gamma}(\lambda_{\alpha}-\lambda^{'}_{\gamma})
         \varepsilon_{\gamma}(\lambda_{\gamma}^{'}|\lambda^{\pm}_{\nu})
                 {\rm d} \lambda_{\gamma}^{'} , \cr
   \sigma_{\infty \alpha}(\lambda_{\alpha}|\lambda^{\pm}_{\nu})
&=&
   \sigma_{\infty \alpha}^{0}(\lambda_{\alpha}) \cr
&+& \sum_{\gamma =1}^{n-1} \int_{\lambda^{-}_{\gamma}}^{\lambda^{+}_{\gamma}}
               K_{\alpha \gamma}(\lambda_{\alpha}-\lambda^{'}_{\gamma})
        \sigma_{\infty \gamma}(\lambda_{\gamma}^{'}|\lambda^{\pm}_{\nu})
                     {\rm d} \lambda_{\gamma}^{'} ,
\nonumber
\end{eqnarray}
with 
$\sigma_{\infty \alpha}^{0}(\lambda_{\alpha}) 
 = \frac{1}{2\pi} \Theta^{'}_{2} (\lambda_{\alpha}) \delta_{\alpha 1} $
and 
$K_{\alpha \gamma}(\lambda)
=\frac{1}{2 \pi} \Theta^{'}_{\alpha \gamma}(\lambda) 
$. 
The quantity 
$\varepsilon_{\alpha}(\lambda_{\alpha}|\lambda^{\pm}_{\nu})$ 
is called the dressed energy,  and 
the "Fermi points" $\lambda_{\alpha}^{\pm}$
of massless spinons are determined by the condition,  
$\varepsilon_{\alpha}(\lambda_{\alpha}^{\pm}|\lambda^{\pm}_{\nu})=0$. 
As should be expected, the expression for the 
ground-state energy (\ref{eqn:su-5}) is typical 
for $(n-1)$ independent $c=1$ Gaussian CFTs, which 
indeed coincides with the known formula 
for the SU($n$) spin chain.\cite{suzuki,izergin} 
So, one cannot see the effect of the holon at this stage.

Let us next analyze the excited 
energy $\Delta E=E-E_{0}$ and the associated 
momentum $\Delta P=P-P_{0}$ when a hole is 
created by the photoemission. 
Because of spin-charge decoupling, the excited 
energy and momentum are naturally 
written as a sum of a holon term and a spinon term, 
\begin{eqnarray}
\Delta E &=& \Delta \varepsilon_{c}(q)
                      +\Delta \varepsilon_{s}(\lambda^{\pm}_{\nu}), \cr
\Delta P &=& \Delta p_{c}(q)
                         +\Delta p_{s}(\lambda^{\pm}_{\nu}),
\label{eqn:su-8}
\end{eqnarray}
where $\Delta \varepsilon_{c}(q)$ ($\Delta p_{c}(q)$) 
is the excitation energy (momentum) of a massive holon 
while $\Delta \varepsilon_{s}(\lambda^{\pm}_{\nu})$ 
($\Delta p_{s}(\lambda^{\pm}_{\nu})$) are those of $(n-1)$ 
types of spinons. 

Since the quantities for holon, 
$\Delta \varepsilon_{c}(q)$  and $\Delta p_{c}(q)$,
can be easily calculated, which give the holon dispersion,
we shall concentrate on  the finite-size corrections 
in the massless spin sector, 
$\Delta \varepsilon_{s}(\lambda^{\pm}_{\nu}), 
\Delta p_{s}(\lambda^{\pm}_{\nu})$, 
by properly treating a final-state interaction. 
The energy for the spin sector is given by 
\begin{eqnarray}
  \varepsilon_{s}(\lambda^{\pm}_{\nu}) =
  N \sum_{\alpha=1}^{n-1} 
   \int_{\lambda^{-}_{\alpha}}^{\lambda^{+}_{\alpha}}
    \sigma_{\alpha}^{0} (\lambda_{\alpha})
    \varepsilon_{\alpha}(\lambda_{\alpha}|\lambda^{\pm}_{\nu}) {\rm d} \lambda_{\alpha},
\label{eqn:su-9}
\end{eqnarray}
where 
$\sigma_{\alpha}^{0} (\lambda_{\alpha})=
 \sigma_{\infty,\alpha}^{0} (\lambda_{\alpha})+
 \frac{1}{N} \sigma_{c,\alpha}^{0} (\lambda_{\alpha})$ 
and 
$\sigma_{c,\alpha}^{0} (\lambda_{\alpha})
 = \frac{1}{2\pi} \Theta_{2}^{'} (\lambda_{\alpha}-q) 
\delta_{\alpha n-1}$.  
Let us expand $\varepsilon_{s}(\lambda^{\pm}_{\nu})$ 
around the ground-state energy 
$N \varepsilon_{0}=\varepsilon_{s}(\pm \lambda^{0}_{\nu})$, 
\begin{eqnarray}
   \Delta \varepsilon_{s}(\lambda^{\pm}_{\nu}) = 
   \frac{1}{2} \sum_{\alpha=1}^{n-1} 
    \left\{
      \frac{\partial^{2} \varepsilon_{s}}{\partial \lambda^{+2}_{\alpha}}
       \left|_{\lambda^{+}_{\alpha}=\lambda^{0}_{\alpha}} \right.
        (\lambda^{+}_{\alpha}-\lambda^{0}_{\alpha})^{2} \right. \cr
\hspace*{\fill}    +
    \left.
      \frac{\partial^{2} \varepsilon_{s}}{\partial \lambda^{-2}_{\alpha}}
       \left|_{\lambda^{-}_{\alpha}=-\lambda^{0}_{\alpha}} \right.
        (\lambda^{-}_{\alpha}+\lambda^{0}_{\alpha})^{2}      
    \right\}. \cr
\label{eqn:su-10}
\end{eqnarray}
Recall here that  the deviation of the Fermi points,
$\lambda^{+}_{\alpha}-\lambda^{0}_{\alpha}$ 
and $\lambda^{-}_{\alpha}+\lambda^{0}_{\alpha}$, 
should be related to the change of the particle number
and also the current induced in the system. 
First, we note that the number of particles $M_\alpha^{0}$ 
and the current $D_\alpha^{0}$
for each spinon in the ground state is given by 
\begin{eqnarray}
      \frac{M_{\alpha}^{0}}{N} &=& 
       \int_{-\lambda_{\alpha}^{0}}^{+\lambda_{\alpha}^{0}}
       \sigma_{\infty \alpha} (\lambda_{\alpha} | \pm \lambda_{\nu}^{0}) , \cr
      \frac{D^{0}_{\alpha}}{N} &=& 
         \frac{1}{2} \left(
                           \frac{1}{2\pi}z_{\infty\alpha} (+\infty)+
                           \frac{1}{2\pi}z_{\infty\alpha} (-\infty)
                       \right)\cr
     &-&\frac{1}{2} \left(
     \int_{+\lambda_{\alpha}^{0}}^{\infty} 
      \sigma _{\infty \alpha} (\lambda_{\alpha} | \pm \lambda_{\nu}^{0}) 
     -
     \int^{-\lambda_{\alpha}^{0}}_{-\infty} 
      \sigma _{\infty \alpha} (\lambda_{\alpha} | \pm \lambda_{\nu}^{0})\right),
\nonumber
\end{eqnarray}
where 
\begin{eqnarray}
     z_{\infty \alpha} (\lambda_{\alpha})
     &=&\Theta_{2} ( \lambda_{\alpha} ) \delta_{\alpha 1} \cr 
     &+&\sum_{\gamma=1}^{n-1} 
             \int_{-\lambda_{\alpha}^{0}}^{+\lambda_{\alpha}^{0}}
    \Theta_{\alpha \gamma} ( \lambda_{\alpha} - \lambda_{\gamma}^{'} )
    \sigma _{\infty \gamma} (\lambda_{\gamma}^{'} | \pm \lambda_{\nu}^{0})
             {\rm d} \lambda_{\gamma}^{'}.
\nonumber
\end{eqnarray}
On the other hand, if one electron is emitted from the system 
the above  relations should read  
\begin{eqnarray}
    \frac{M^{0}_{\alpha}+1}{N} 
         &=& \int_{\lambda_{\alpha}^{-}}^{\lambda_{\alpha}^{+}}
              \sigma_{\alpha} (\lambda_{\alpha} | \lambda_{\nu}^{\pm}) \cr
         &=& \int_{\lambda_{\alpha}^{-}}^{\lambda_{\alpha}^{+}} 
  \left(
     \sigma_{\infty \alpha}(\lambda_{\alpha}|\lambda^{\pm}_{\nu})
    + \frac{1}{N}\sigma_{c,\alpha}(\lambda_{\alpha}|\lambda^{\pm}_{\nu})\right),\cr
    \frac{D^{0}_{\alpha}+1/2\delta_{\alpha,1}}{N}
         &=& \frac{1}{2} \left(
                             \frac{1}{2\pi}z_{\alpha} (+\infty)+
                             \frac{1}{2\pi}z_{\alpha} (-\infty)
                       \right)\cr
     &-&\frac{1}{2} \left(
        \int_{\lambda_{\alpha}^{+}}^{\infty} 
        \sigma _{\alpha} (\lambda_{\alpha} | \lambda_{\nu}^{\pm}) 
     -
        \int^{\lambda_{\alpha}^{-}}_{-\infty} 
        \sigma _{\alpha} (\lambda_{\alpha} | \lambda_{\nu}^{\pm})\right),
\nonumber
\end{eqnarray}
where 
\begin{eqnarray}
z_{\alpha} (\lambda_{\alpha})
     &=&\Theta_{2} ( \lambda_{\alpha} ) \delta_{\alpha 1} +
         \frac{1}{N} \Theta_{2} ( \lambda_{\alpha} -q) \delta_{\alpha n-1} \cr
     &+& \sum_{\gamma=1}^{n-1} \int_{\lambda_{\alpha}^{-}}^{\lambda_{\alpha}^{+}}
        \Theta_{\alpha \gamma} ( \lambda_{\alpha} - \lambda_{\gamma}^{'} )
          \sigma _{\gamma} (\lambda_{\gamma}^{'} | \lambda_{\nu}^{\pm}) 
                 {\rm d} \lambda_{\gamma}^{'}.
\nonumber
\end{eqnarray}
Therefore, one can find that the change in the particle number 
$\Delta M_{\alpha}=M_{\alpha}-M_{\alpha}^{0}$  
and the induced current $\Delta D_{\alpha}=D_{\alpha}-D_{\alpha}^{0}$ 
satisfy  
\begin{eqnarray}
   \frac{\Delta M_{\alpha}+1-n_{c,\alpha}}{N} 
    &=&
   \sum_{\beta =1}^{n-1} 
       \xi_{\alpha \beta}(\lambda_{\beta}^{0} )
     \sigma _{\infty\beta} (\lambda_{\beta}^{0} )
        ( \lambda_{\beta}^{+} - \lambda_{\beta}^{0} )\cr
    &-&
        \xi_{\alpha \beta}(-\lambda_{\beta}^{0} )
    \sigma _{\infty\beta} (-\lambda_{\beta}^{0} )
           ( \lambda_{\beta}^{-} + \lambda_{\beta}^{0} ),\cr
   \frac{\Delta D_{\alpha}+1/2\delta_{\alpha,1}-d_{c,\alpha}}{N}
      &=&
    \sum_{\beta=1}^{n-1}  
    z_{\alpha \beta}(\lambda_{\beta}^{0} )
    \sigma _{\infty\beta} (\lambda_{\beta}^{0} )
    ( \lambda_{\beta}^{+} - \lambda_{\beta}^{0} )\cr
      &+&
    z_{\alpha \beta}(-\lambda_{\beta}^{0} )
    \sigma _{\infty\beta} (-\lambda_{\beta}^{0} )
    ( \lambda_{\beta}^{-} + \lambda_{\beta}^{0} ),\cr
&\hspace{1mm}&
\label{eqn:su-1.3}
\end{eqnarray}
where we have  neglected terms of $O(1/N^{2})$.
We can see from (\ref{eqn:su-3}) 
that the quantum numbers 
$\Delta M_{\alpha}$ and $\Delta D_{\alpha}$ 
are subject to the selection rule (gluing condition)
\begin{eqnarray}
    \Delta D_{\mu} = -\frac{1}{2} (\Delta M_{\mu -1} 
                                 +\Delta M_{\mu +1})
                               +\Delta M_{\mu +1} \delta_{\mu 1} 
                          \hspace{2mm} {\rm mod} \hspace{2mm}1.\cr
\label{eqn:su-14}
\end{eqnarray}
 The quantities 
$\xi_{\alpha \beta}=\xi_{\alpha \beta}(\lambda_{\beta}^{0} | \pm \lambda^{0}_{\nu})$ 
are the so called dressed charges which are given by the solution to 
the integral equations\cite{izergin} 
\begin{eqnarray}
   \xi_{\alpha \beta}(\lambda_{\beta} | \pm \lambda^{0}_{\nu})
&=& \delta_{\alpha \beta} \cr
&+& \sum_{\gamma =1}^{n-1} 
                \int_{-\lambda^{0}_{\gamma}}^{+\lambda^{0}_{\gamma}}
       \xi_{\alpha \gamma}(\lambda_{\gamma}^{'} | \pm \lambda^{0}_{\nu})
                 K_{\gamma \beta}(\lambda_{\gamma}^{'}-\lambda_{\beta})
                         {\rm d} \lambda_{\gamma}^{'}.
\nonumber
\end{eqnarray}
Note that the quantities of 
$z_{\alpha \beta}(\lambda_{\beta}^{0} | \pm \lambda^{0}_{\nu})$ 
introduced in (\ref{eqn:su-1.3})  
are related to the dressed charges:
$2\sum_{\beta}z_{\alpha \beta}
\xi_{\beta \gamma}=\delta_{\alpha \gamma}$. 

A remarkable point in  (\ref{eqn:su-1.3}) is that  
two kinds of phase shifts $n_{c,\alpha}$ and $d_{c,\alpha}$, 
 which are the key quantities to
control the anomalous low-energy properties,
enter in (\ref{eqn:su-1.3}).
These phase shifts, which are caused by 
a final-state interaction between spinons and
the created holon, are explicitly obtained as
\begin{eqnarray}
   n_{c,\alpha} &=& \int_{-\lambda_{\alpha}^{0}}^{+\lambda_{\alpha}^{0}}
                      \sigma_{c,\alpha} (\lambda_{\alpha} | 
                                    \pm \lambda_{\nu}^{0}) , \cr
   d_{c,\alpha}&=&\frac{1}{2}
                \left( 
                  \frac{1}{2\pi} z_{c,\alpha}(\infty)
                 +\frac{1}{2\pi} z_{c,\alpha}(-\infty)  
                \right)\cr
\vspace{2mm}
               &-&\frac{1}{2}
                \left(
                 \int_{+\lambda_{\alpha}^{0}}^{\infty} 
                 \sigma _{c,\alpha} (\lambda_{\alpha} | 
                         \pm \lambda_{\nu}^{0}) 
                -
                 \int^{-\lambda_{\alpha}^{0}}_{-\infty} 
                 \sigma _{c,\alpha} (\lambda_{\alpha} | 
                           \pm \lambda_{\nu}^{0})\right),
\nonumber
\end{eqnarray}
where 
\begin{eqnarray}
   \sigma_{c,\alpha}(\lambda_{\alpha}|\lambda^{\pm}_{\nu})
    &=&
   \sigma_{c,\alpha}^{0}(\lambda_{\alpha}) \cr
    &+& \sum_{\gamma =1}^{n-1} 
           \int_{-\lambda^{0}_{\gamma}}^{+\lambda^{0}_{\gamma}}
            K_{\alpha \gamma}(\lambda_{\alpha}-\lambda^{'}_{\gamma})
            \sigma_{c,\gamma}(\lambda_{\gamma}^{'}|\pm \lambda_{\nu})
                       {\rm d} \lambda_{\gamma}^{'}, \cr
   z_{c,\alpha} (\lambda_{\alpha})
    &=&
          \Theta_{2} ( \lambda_{\alpha} -q) \delta_{\alpha n-1} \cr
    &+&\sum_{\gamma=1}^{n-1} 
                 \int_{-\lambda_{\alpha}^{0}}^{+\lambda_{\alpha}^{0}}
    \Theta_{\alpha \gamma} ( \lambda_{\alpha} - \lambda_{\gamma}^{'} )
      \sigma _{c,\gamma} (\lambda_{\gamma}^{'} | \pm \lambda_{\nu}^{0}) 
                 {\rm d} \lambda_{\gamma}^{'}.
\nonumber
\end{eqnarray}
Combining the above expressions, we end up with
the excitation energy,
\begin{eqnarray}
 \Delta \varepsilon_{s}=\sum_{\alpha=1}^{n-1} \frac{2\pi v_{\alpha}}{N}
   \{
    \frac{1}{4} ({\bf\xi}^{-1} {\bf n})_{\alpha}^{2} +
    ({\bf \xi}^{T} {\bf d})_{\alpha}^{2} +
    N_{\alpha}^{+}+ N_{\alpha}^{-}
   \}, 
\nonumber
\end{eqnarray}
where we have added particle-hole excitations
specified by the quantum numbers $N_{\alpha}^{+}$ and $N_{\alpha}^{-}$. 
Here we have introduced the notation 
\begin{eqnarray}
  ({\bf\xi})_{\alpha \beta}=\xi_{\alpha \beta},\hspace{2mm}
  ({\bf n})_{\alpha}&=&\Delta M_{\alpha}+1-n_{c,\alpha},\hspace{2mm} \cr
  ({\bf d})_{\alpha}&=&\Delta D_{\alpha}+\frac{1}{2}\delta_{\alpha,1} -d_{c,\alpha}.
\end{eqnarray} 

Performing a similar manipulation, the excited momentum 
$\Delta p_{s}$ is also obtained as 
\begin{eqnarray}
   \Delta p_{s}&=& 2\pi \sum_{\alpha=1}^{n-1} Q_{F,\alpha} ({\bf d})_{\alpha}\cr
               &+&\frac{2\pi}{N}\sum_{\alpha=1}^{n-1} 
   \{
     ({\bf n})_{\alpha} +
     ({\bf d})_{\alpha} +
    N_{\alpha}^{+}- N_{\alpha}^{-}
   \}.
\label{eqn:su-18}
\end{eqnarray} 
This completes the calculation of the 
finite-size corrections.  

\subsection{Conformal dimensions}

According to the finite-size scaling in CFT, 
conformal dimensions $\Delta_{\alpha}^{\pm}$ 
of $\alpha$-type spinons are read from 
the universal $1/N$ corrections to the excitation 
energy $\Delta \varepsilon_{s}$ 
and momentum $\Delta p_{s}$, which are given by 
\begin{eqnarray}
  \Delta _{\alpha}^{+}+\Delta _{\alpha}^{-}
   &=&
  \frac{1}{4} ({\bf\xi}^{-1} {\bf n})_{\alpha}^{2} +
    ({\bf\xi}^{T} {\bf d})_{\alpha}^{2} +
    N_{\alpha}^{+}+ N_{\alpha}^{-}, \cr
\vspace{2mm}
  \Delta _{\alpha}^{+}-\Delta _{\alpha}^{-}
   &=&
  ({\bf n})_{\alpha} +
     ({\bf d})_{\alpha} +
    N_{\alpha}^{+}- N_{\alpha}^{-}.
\label{eqn:su-19}
\end{eqnarray}
So, they are reduced to 
\begin{eqnarray}
   \Delta _{\alpha}^{\pm}
    =
   \frac{1}{2}
    \left(
     \frac{1}{2} {\bf\xi}^{-1} {\bf n} \pm {\bf\xi}^{T} {\bf d}\right)^{2}_{\alpha}
     \pm N_{\alpha}^{\pm}.
\label{eqn:su-20}
\end{eqnarray}
Although conformal dimensions $\Delta_{\alpha}^{\pm}$ 
for massless spinons are typical for 
$c=1$ CFTs,\cite{fk,kawakami} the massive holon 
also contributes to $\Delta_{\alpha}^{\pm}$ via 
the phase shifts $n_{c,\alpha}(q)$ and $d_{c,\alpha}(q)$.
In this sense, (\ref{eqn:su-20}) is classified as 
{\it shifted} $c=1$ CFTs, whose fixed point is different 
even from that of the static 
impurity problem, as pointed out by Sorella and Parola.\cite{S&P2} 
This fixed point indeed belongs to that  of 
{\it mobile-impurity class} in 1D quantum systems,\cite{tsukamoto}
for which two kinds of phase shifts play a crucial role.

\section{One-particle Green function and Photoemission Spectra}

\subsection{Critical properties for holon dispersion}\label{subsec:3-1}
Having classified low-energy properties by CFT,
we are now ready to investigate the critical properties of 
the one-electron Green function  at absolute zero, 
which is defined by  
\begin{eqnarray}
   && G_{\beta}(x,t)={\rm i} \sum_{q} \theta(-t)
               {\rm e}^{{\rm i}\Delta \varepsilon_{c}t-{\rm i}\Delta p_{c}x} \cr
               &\times&
               \sum_{N,\bar{N}=0}^{\infty}
               <0|\psi^{\dagger}(0)|N,\bar{N}><N,\bar{N}|\psi(0)|0>
               {\rm e}^{{\rm i}\Delta \varepsilon_{s}t-{\rm i}\Delta p_{s}x}, \cr
&\hspace*{1mm}&\label{eqn:su-21}
\end{eqnarray}
for $\beta = 0,\cdots,n-1$, where 
$|0>$ represents the ground state.
Exploiting finite-size scaling in CFT, we can 
write down its asymptotic form as
\begin{eqnarray}
   \sum_{N,\bar{N}=0}^{\infty}
               <0|\psi^{\dagger}(0)|N,\bar{N}><N,\bar{N}|\psi(0)|0>
               {\rm e}^{{\rm i}\Delta \varepsilon_{s}t-{\rm i}\Delta p_{s}x} \cr
   \rightarrow \prod_{\alpha=1}^{n-1}
   \frac{e^{  
            -i 2\pi Q_{F,\alpha} ({\bf d})_{\alpha} x 
           }
        }
        {
        (x-v_{\alpha}t)^{2 \Delta _{\alpha}^{+}} 
        (x+v_{\alpha}t)^{2 \Delta _{\alpha}^{-}}
        },
\nonumber
\label{eqn:su-22}
\end{eqnarray}
where $\Delta^{\pm}_{\alpha}$ are conformal dimensions 
which are related to the scaling dimensions $x_{\beta}$, 
\begin{eqnarray}
   x_{\beta}&=&\sum_{\alpha=1}^{n-1}
       (\Delta _{\alpha}^{+}+\Delta _{\alpha}^{-}) \cr
    &=&
   \frac{1}{4} 
    {\bf n}^{T} ({\bf\xi}^{-1})^{T} ({\bf \xi}^{-1}) {\bf n} + 
    {\bf d}^{T}  {\bf\xi} {\bf \xi}^{T} {\bf d}.
\label{eqn:su-23}
\end{eqnarray}
The asymptotic form of the spectral function $A_{\beta}(k,w)$ 
around the holon dispersion is 
given via the Fourier transformation as 
\begin{eqnarray}
  A_{\beta} (k,\omega) = \frac{1}{\pi} {\rm Im} G_{\beta}(k,\omega)
                     \sim ( \omega - \omega_{c}(k-Q) )^{X_{\beta}(k)} \cr
\label{eqn:su-2.3}
\end{eqnarray}
with the critical exponent 
\begin{eqnarray}
X_{\beta}(k)=2x_{\beta}-1, 
\end{eqnarray}
where the energy
$\omega_{c}(k-Q)= \Delta \varepsilon_{c}(q)$ and 
the momentum 
$Q=2\pi \sum_{\alpha=1}^{n-1}Q_{F,\alpha} ({\bf d})_{\alpha}$
feature the dispersion of holon. 
We can see that the singularity in the 
spectral function which occurs at 
frequencies determined by the holon dispersion is governed   
by the critical exponent $X_{\beta}(k)$ for spinons.\cite{S&P2} 
Namely, this exponent reflects the infrared divergence
properties of spinons, which also 
includes  the phase shifts  $n_{c,\alpha},d_{c,\alpha}$ 
 caused by a final state interaction.
These two phase shifts usually depend on a rapidity 
of the holon excitation, giving rise  to the momentum dependent  
critical exponents in generic cases.

Now our problem is to read the correct scaling dimensions
by appropriately choosing a set of quantum numbers,
to describe the critical behavior of photoemission 
spectrum.  The key to choose the quantum numbers is the
selection rule given in (\ref{eqn:su-14}).
Since we are now considering the situation that one electron is 
emitted from the system, one holon and one spinon 
should be removed.  This gives the guideline to 
set the quantum numbers. 
Suppose that an electron of $\beta$-type is
emitted from the system,
then we rewrite the quantum numbers as, 
$\Delta M_{\alpha} \rightarrow (\Delta {\bf M}^{\beta})_{\alpha}$ 
 and
$\Delta D_{\alpha} \rightarrow (\Delta {\bf D}^{\beta})_{\alpha}$, 
which can be  chosen as 
\begin{eqnarray}
 (\Delta {\bf M}^{\beta})_{\alpha}&=&-1\hspace{2mm}(0 \leq \alpha \leq \beta), 
               \hspace{2mm}0\hspace{2mm}(\beta < \alpha \leq n-1), \cr
 (\Delta {\bf D}^{\beta})_{\alpha}&=&-\frac{1}{2} \delta_{\alpha 1}
                                     +\frac{1}{2} \delta_{\alpha \beta}
                                     -\frac{1}{2} \delta_{\alpha \beta+1}. 
\label{eqn:su-2.1}
\end{eqnarray}

Let us first consider the SU($n$) case without orbital splitting.
In this case, the $(n-1)\times(n-1)$
dressed charge matrix is easily evaluated as,
\begin{eqnarray}
  ({\bf \xi}^{-1})^{T} ({\bf \xi}^{-1})
   =
\left(
\matrix {2  & -1      & \null  & \null   \cr
         -1         & 2       &  -1    & \null   \cr
        \null       & \ddots  & \ddots & \ddots  \cr
        \null       & \null   &  -1    & 2       \cr}\right), 
\end{eqnarray}
which is nothing but the Cartan matrix for SU($n$)
Lie algebra.\cite{suzuki}
Note that $ {\bf \xi} {\bf \xi}^{T}$ is given by the 
inverse of the above matrix.
By substituting the quantum numbers, we consequently obtain the 
corresponding scaling dimension,
\begin{eqnarray}
x_{\beta}(k)= \frac{1}{2}(1-\frac{1}{n})
\end{eqnarray}
for $\beta=0,1, \cdots, n-1$.
Thus the critical exponent for the holon dispersion is 
$X_{\beta}=-1/n$. Therefore, if the orbital degeneracy becomes 
large, the singular property is expected to become weaker.
In the special case of $n=2$,
this exponent was already obtained.\cite{S&P2,voit}
Note that the above exponent depends neither on the 
spin and orbital indices nor on the momentum.
This simple result follows from that  
SU($n$) symmetry holds exactly in the 
case  without orbital splitting.

We now wish to discuss the effect of orbital
splitting on the critical exponents. In the following,
we concentrate on the specific case of two orbital ($n=2L=4$)
 to see our discussions more clearly,\cite{itakura} since
the generalization to SU($n$)  can be done straightforwardly.
We note that such an orbital splitting may come from the 
crystalline field effects.
In the SU(4) case, there appear three types of spinons, the 
Fermi points of which are:  
$\lambda_{2}^{0}$ is finite, which is due to U(1)
symmetry induced by an orbital splitting
$\Delta_{1}=\Delta$,  
while $\lambda_{1}^{0}$ and $\lambda_{3}^{0}$ are infinite,
reflecting that each spinon excitation in the same orbital 
still has SU(2) symmetry.
Accordingly,  we have the matrix,\cite{itakura}
\begin{eqnarray}
  ({\bf \xi}^{-1})^{T} ({\bf \xi}^{-1})
   &=&
  \left( \begin{array}{ccc}
        2   &  -1                          &  0 \cr
        -1   & 1+\frac{1}{\xi_{\Delta}^{2}} & -1 \cr
        0   &  -1                          &  2 
         \end{array}\right)
,\hspace{2mm} \cr
   {\bf \xi} {\bf \xi}^{T}
   &=&
  \left( \begin{array}{ccc}
1+\frac{\xi_{\Delta}^{2}}{2}  &  \xi_{\Delta}^{2} & \frac{\xi_{\Delta}^{2}}{2}\cr
\xi_{\Delta}^{2}              & 2\xi_{\Delta}^{2} & \xi_{\Delta}^{2}\cr
\frac{\xi_{\Delta}^{2}}{2}    &  \xi_{\Delta}^{2} & 1+\frac{\xi_{\Delta}^{2}}{2}
         \end{array}\right).
\label{eqn:su-24}
\end{eqnarray}
The dressed charge $\xi_{\Delta} \equiv \xi_{\Delta}(\lambda_{2}^{0})$ 
for U(1) part is given by the integral equation, 
\begin{eqnarray}
  \xi_{\Delta}(\lambda|\pm \lambda_{2}^{0})
=
  1+\int_{-\lambda_{2}^{0}}^{+\lambda_{2}^{0}}
         G(\lambda - \lambda^{'})
         \xi_{\Delta}(\lambda^{'}|\pm \lambda_{2}^{0}),
\end{eqnarray}
where the integral kernel $G(\lambda)$ is
\begin{eqnarray}
G(\lambda)=\int_{-\infty}^{\infty}
               \frac{{\rm d}k}{2\pi}
               \frac{1-{\rm e}^{|k|}}{1+{\rm e}^{|k|}} {\rm e}^{i k \lambda}.
\end{eqnarray}
As mentioned above, this system has SU(4) symmetry in  
a vanishing orbital-splitting, 
so that $\xi_{\Delta} \rightarrow 1$ and 
$({\bf \xi}^{-1})^{T} ({\bf \xi}^{-1})$ becomes 
SU(4) Cartan matrix.

We rewrite the scaling dimensions by using (\ref{eqn:su-24}), 
\begin{eqnarray}
 x_{0}&=& x_{1}=
\frac{1}{4} + \frac{1}{4\xi_{\Delta}^{2}} (1-n_{c})^{2}+
                      \xi_{\Delta}^{2} (-\frac{1}{4}-d_{c})^{2}, \cr
 x_{2}&=&x_{3}= \frac{1}{4} + \frac{1}{4\xi_{\Delta}^{2}} (-n_{c})^{2}+
                        \xi_{\Delta}^{2} (\frac{1}{4}-d_{c})^{2}, 
\label{eqn:su-25}
\end{eqnarray}
where,
\begin{eqnarray}
   n_{c}&=&\int^{+\lambda_{2}^{0}}_{-\lambda_{2}^{0}}
         \sigma_{c}(\lambda|\pm \lambda_{2}^{0})  \cr
   d_{c}&=&\frac{1}{2}
         \left( \int^{-\lambda_{2}^{0}}_{-\infty}
                 \sigma_{c}(\lambda|\pm \lambda_{2}^{0}) 
                -
                \int^{\infty}_{+\lambda_{2}^{0}}
                 \sigma_{c}(\lambda|\pm \lambda_{2}^{0})\right),
\nonumber
\end{eqnarray}
and 
\begin{eqnarray}
\sigma_{c}(\lambda|\pm \lambda_{2}^{0})
 &=&
  \sigma_{c}^{0}(\lambda)+
    \int^{+\lambda_{2}^{0}}_{-\lambda_{2}^{0}}
    G(\lambda-\lambda^{'})
    \sigma_{c}(\lambda^{'}|\pm \lambda_{2}^{0}), \cr
\vspace{2mm}
  \sigma_{c}^{0}(\lambda)
 &=& 
  \int^{\infty}_{\infty}
   \frac{{\rm d}k}{2\pi}
   \frac{{\rm e}^{-|k|/2}}{1+{\rm e}^{|k|}}
   {\rm e}^{ik(\lambda-q)}.
\end{eqnarray}
Here $x_{0,1}$ ($x_{2,3}$) are the scaling dimensions
for the case that one electron is emitted from lower-orbital 
(upper-orbital) band electrons.
Note that the scaling dimensions  depend on 
the momentum of the created  hole through the 
phase shifts $n_{c}$ and $d_{c}$. 
One can also see that both of the scaling 
dimensions in (\ref{eqn:su-25})
include the term $\frac{1}{4}$ which comes from  
level-1 SU(2) CFT for spin excitations in the same orbital.
On the other hand, orbital excitation is described by U(1) CFT
which is featured by the dressed charge $\xi_{\Delta}$,
since level-1 SU(4) CFT is reduced to two level-1 SU(2) CFTs and U(1)
CFT in the presence of  orbital splitting.

The critical exponent $X_{\beta}(k)$ is shown in Fig. 1, where  
the momentum is plotted in the unit of inverse lattice 
spacing $1/a$.
In Fig. 1(a), the critical exponent is shown for the case
of an electron being emitted from the upper-orbital band
for several choices of the orbital splitting. It is seen that
the critical exponent is strongly dependent on
the momentum. In the range where the critical exponent 
becomes close to zero, the divergence singularity is
weakened, whereas for larger negative values of the exponent, 
the singularity becomes stronger.
Shown in Fig. 1(b) is the critical exponent when an 
electron is emitted from the lower-orbital band.  
In this case also, the exponent is strongly momentum-dependent,
and it can change even its sign, which means that divergent
power-law singularity is changed into convergent power-law property
in some momentum range.
We note that when the orbital splitting $\Delta$ is larger 
than a critical value $\Delta_{c}$, 
the upper orbital becomes empty without particles. 
Then the relevant effective theory simply becomes
level-1 SU(2) CFT, and the scaling dimension becomes $\frac{1}{4}$ 
which was previously obtained by Sorella and Parola,\cite{S&P1,S&P2} 
and Voit.\cite{voit} 

%
\subsection{Critical properties for spinon dispersions}

We have so far discussed the critical behavior 
for the holon dispersion.
In the photoemission experiments, 
the large spectral intensity has been observed 
not only for the holon dispersion but also 
for the spinon dispersion. 
In this section we study critical behavior of 
photoemission spectra around the spinon dispersions.
This problem was recently addressed by Voit by using the 
Luther-Emery model which does not 
include the effect of  orbital degeneracy.\cite{voit}
His calculation was based on several conjectures 
for the spectral functions, because 
even for the Luther-Emery model the exact dynamical 
correlation functions cannot be obtained. We shall confirm 
the validity of his conjectures 
by using the CFT analysis, and also extend his results to 
the multi-orbital case.

Let us first recall that in the case of $J\rightarrow 0$
(or $U \rightarrow \infty$ Hubbard model),
the spectra on the spinon dispersion is determined by a band edge 
singularity of holon.\cite{S&P1,nagaosa}
We shall see, however,
that  in the case of $J \neq 0$ spinon excitations at the 
low-energy regime becomes essential, which indeed  
controls the critical behaviour of spectral functions on the spinon 
dispersions. In order to 
observe critical properties of spectra on the spinon dispersions, 
we should consider a massive holon with 
the lowest excitation energy (rapidity $q=0$), 
which  gives the energy shift for 
the spinon dispersion. 
Also, as mentioned before, the holon behaves as if it is  
a mobile impurity, and affects the critical properties 
of spinons through the phase shifts caused by 
a final state interaction.

Since the SU($n$) spin model has
 $(n-1)-$types of spinons (this number corresponds to 
the rank of the underlying SU($n$) Lie algebra), 
there show up $(n-1)$ spinon dispersions in the low-energy regime. 
Though it is not easy to perform  Fourier transformation of 
(\ref{eqn:su-21}) completely, we can write down its asymptotic form
around each spinon dispersion, by 
extending the results in ref. 17 
where a metallic electron system was studied.  
The spectral function $A_{\beta a}(k,w)$ on each spinon dispersion 
specified by suffix $a$ is given by 
\begin{eqnarray}
A_{\beta a}(k,\omega) \sim 
(\tilde{\omega}-v_{a} \tilde{k})^{X_{\beta a}},
\label{eqn:su-26}
\end{eqnarray}
where $\tilde{\omega}=\omega - \Delta \epsilon_{c}(q=0) $ and 
$\tilde{k}=k-\Delta p_{c}(q=0) $. 
One can see that the energy is shifted by 
the amount of the lowest excitation energy of 
holon, $\Delta \epsilon_{c}(q=0)$.
We find that the corresponding critical 
exponents $X_{\beta a}$ are obtained in terms of a 
specific combination of conformal dimensions, 
\begin{eqnarray}
  X_{\beta a}=2 \Delta_{a}^{-} +
        \sum_{i \neq a}^{n-1} (2\Delta_{i}^{+} + 2\Delta_{i}^{-}) -1
\end{eqnarray}
for $a=1,\cdots,n-1$,  where 
$\Delta_{\alpha}^{\pm}$ are conformal dimensions 
which are obtained in (\ref{eqn:su-20}). 
We should notice again that  (\ref{eqn:su-26}) is not the 
ordinary one-particle Green function of spinons, 
but indeed  includes the effect of a massive holon
via the phase shifts. 

We first consider the SU($n$) symmetric case without
orbital splitting.  As mentioned above, the critical exponent
does not depend on the momentum in this case
because of high spin-orbital symmetry of the system, 
and $X_{\beta a}$ takes $-1/n$. In the case 
of SU(2), $X_{\beta a}=-1/2$, which agrees with the value
previously obtained.\cite{voit}  Let us next discuss 
the effect of orbital splitting by using the two-band model 
introduced in \ref{subsec:3-1}. 
In the low-frequency regime we are now interested in, 
$X_{\beta a}$ is independent of the momentum even in the 
presence of the splitting, so that 
we show the values of $X_{\beta a}$ as a 
function of the orbital splitting in Fig. 2. 
In Figs. 2(a) and 2(b) the critical exponents are shown for the case of 
one electron being emitted from the upper and the lower band,
respectively. Since each  band has SU(2) spin symmetry, 
$X_{\beta 1}=X_{\beta 3}$. 
When the orbital splitting goes to zero, all the exponents 
approach the same value $-1/4$ characteristic of SU(4)
case. It is seen from Fig. 2(b) that the exponent can change 
its sign as the increase of orbital splitting,
because of the effect of a final state interaction. 
It is thus seen that either the divergent or convergent 
power-law singularity emerges on each spinon dispersion, 
depending on the value of the orbital splitting.

\section{Summary}
We have studied the critical properties 
in photoemission spectra for the Mott insulator with orbital degeneracy. 
We have calculated the spectral functions and their critical 
exponents for the supersymmetric multi-orbital {\it t-J} model,  
by combining the Bethe ansatz  with finite-size scaling methods 
in CFT. It has been confirmed that
power-law  singularities around the massive holon dispersion 
as well as the ($n-1$) massless spinon dispersions are governed by
the infrared-divergence properties of massless spinons. 
This is partly in contrast to the results for 
the $U \rightarrow \infty$ 
Hubbard model in which a singularity around the spinon dispersion 
reflects  the band-edge structure of the holon dispersion.
It has been clarified  that for the critical behavior in 
$(n-1)$ spinon dispersions, the 
specific combination of conformal dimensions determines 
the corresponding critical exponents, which are generally 
different from that for the holon dispersion.
The effect of the orbital-splitting has  been studied,
which induces the momentum-dependent
critical exponents in the spectral function.

In this paper, we have studied the specific {\it t-J} model 
with supersymmetry ($t=J$). Nevertheless, the conclusions 
obtained in the present paper can be directly applied to 
more general cases ($t \neq J$).  This is because
in the problem of the photoemission, we are concerned with the 
one hole doped in the Mott insulator, so that 
the change in $t$ is expected to merely modify  
the kinetic energy of the doped hole.
Also, we have not considered the effect of the 
Hund coupling. This effect may be particularly 
important to treat the systems with double exchange interaction
(or ferromagnetic Kondo lattice systems). This  
problem is now under consideration.

\section*{Acknowledgements}
This work was partly supported by a Grant-in-Aid from the Ministry
of Education, Science, Sport and Culture, Japan.

%
%
\begin{figure}[h]
\caption{
Momentum-dependent critical exponents $X_{\beta}(k)$ for 
the holon dispersion: (a) and (b) correspond to the cases
when an electron is emitted from 
the upper- and lower-orbital band. }
\label{fig:1}
\end{figure}
\begin{figure}[h]
\caption{
Critical exponents $X_{\beta a}$ for the spinon dispersions: 
(a) and (b) correspond to the cases when 
an electron is emitted from 
the upper- and lower-orbital band.  Note that 
the orbital-splitting is normalized by the exchange 
coupling $J$. 
}
\label{fig:2}
\end{figure}
\end{document}